\documentclass[twocolumn,prl]{revtex4-1}
\usepackage[dvipdfmx]{graphicx}
\usepackage{amssymb}
\usepackage{amsmath}
\usepackage{bm}
\usepackage{color}
\usepackage{17657}

\begin{document}

\title{Quantum Oscillations of Gilbert Damping in  Ferromagnetic/Graphene Bilayer Systems}
\author{Yuya Ominato${}^1$ and Mamoru Matsuo${}^{1,2}$}
\affiliation{${}^1$Kavli Institute for Theoretical Sciences, University of Chinese Academy of Sciences, Beijing 100190, China}
\affiliation{${}^2$CAS Center for Excellence in Topological Quantum Computation, University of Chinese Academy of Sciences, Beijing 100190, China}
\date{\today}

\begin{abstract}
We study the spin dynamics of a ferromagnetic insulator on which graphene is placed. We show that the Gilbert damping is enhanced by the proximity exchange coupling at the interface. The modulation of the Gilbert damping constant is proportional to the product of the spin-up and spin-down densities of states of graphene. Consequently, the Gilbert damping constant in a strong magnetic field oscillates as a function of the external magnetic field that originates from the Landau level structure of graphene. We find that a measurement of the oscillation period enables the strength of the exchange coupling constant to be determined. The results theoretically demonstrate that the ferromagnetic resonance measurements may be used to detect the spin resolved electronic structure of the adjacent materials, which is critically important for future spin device evaluations.
\end{abstract}
\maketitle

{\it Introduction}.---Graphene spintronics is an emergent field aiming at exploiting exotic spin-dependent properties of graphene for spintronics devices \cite{hanGrapheneSpintronics2014}.
Although pristine graphene is a non-magnetic material, there have been efforts to introduce magnetism into graphene to find spin-dependent phenomena and to exploit its spin degrees of freedom.
Placing graphene on a magnetic substrate is a reasonable way, which leads to magnetic proximity effect and lifting of spin degeneracy \cite{yangProximityEffectsInduced2013,hallalTailoringMagneticInsulator2017}.
Subsequently, magnetization was induced in graphene and spin dependent phenomena, such as the anomalous Hall effect \cite{wangProximityInducedFerromagnetismGraphene2015a,averyanovHighTemperatureMagnetismGraphene2018} and non-local spin transport \cite{leutenantsmeyerProximityInducedRoom2016,weiStrongInterfacialExchange2016a}, were observed.
In all these experiments, a spin-dependent current was generated by an electric field. There is an alternative way to generate a spin current called spin pumping \cite{mizukamiStudyFerromagneticResonance2001,mizukamiFerromagneticResonanceLinewidth2001,mizukamiEffectSpinDiffusion2002,rahimiElectricallyControllableSpin2015,inoueSpinPumpingTwodimensional2016}.
The proximity exchange coupling describes spin transfer at the magnetic interface and a spin current is injected using ferromagnetic resonance (FMR) from ferromagnetic materials into the adjacent materials. The generation of a spin current is experimentally detectable through both the inverse spin Hall effect and modulation of the FMR, which were experimentally confirmed at magnetic interfaces between graphene and several magnetic materials \cite{patraDynamicSpinInjection2012,tangDynamicallyGeneratedPure2013,dushenkoGateTunableSpinChargeConversion2016,indoleseWidebandOnChipExcitation2018,mendesSpinCurrentChargeCurrentConversion2015,mendesDirectDetectionInduced2019}.

The theory of spin transport phenomena at magnetic interfaces has been formulated based on the Schwinger-Keldysh formalism \cite{rammerQuantumFieldtheoreticalMethods1986}, which is applicable to magnetic interfaces composed of a variety of systems, such as a paramagnetic metal and a ferromagnetic insulator (FI) \cite{adachiTheorySpinSeebeck2013,ohnumaEnhancedDcSpin2014,ohnumaTheorySpinPeltier2017,matsuoSpinCurrentNoise2018},
a superconductor and FI \cite{inoueSpinPumpingSuperconductors2017,katoMicroscopicTheorySpin2019}, and two FIs \cite{nakataMagnonicNoiseWiedemannFranz2018,nakataAsymmetricQuantumShot2019}.
The modulation of FMR has been investigated in several papers. The modulation of Gilbert damping was found to be proportional to the imaginary part of the dynamical spin susceptibility \cite{simanekGilbertDampingMagnetic2003,ohnumaEnhancedDcSpin2014,tataraConsistentMicroscopicAnalysis2017,inoueSpinPumpingSuperconductors2017,matsuoSpinCurrentNoise2018,katoMicroscopicTheorySpin2019}, which means that one can detect spin excitations and electronic properties of adjacent materials through the FMR measurements. This implies that the FMR measurements of FI/graphene bilayer systems allow us to access the spin-dependent properties of graphene in quantum Hall regime \cite{novoselovTwodimensionalGasMassless2005,zhangExperimentalObservationQuantum2005}. However, the modulation of FMR at the magnetic interface between a FI and graphene has not been investigated and the effect of Landau quantization on the FMR signal is unclear.

\begin{figure*}[t]
\begin{center}
\includegraphics[width=1\hsize]{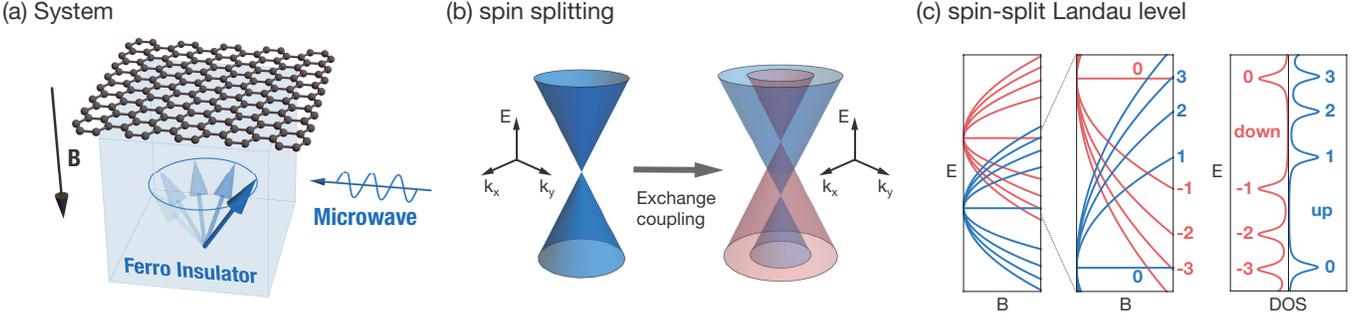}
\end{center}
\caption{
(Color online) Schematic picture of the FMR measurement and the energy spectrum of graphene in a strong perpendicular magnetic field. (a) Graphene on a ferromagnetic insulator substrate. The magnetic field perpendicular to graphene is applied and the microwave is irradiated to the FI. (b) The spin degeneracy is lifted by the exchange coupling. (c) The perpendicular magnetic field leads to the spin-split Landau level structure. The density of states has a peak structure and the level broadening originating from disorder is included.
}
\label{fig_system}
\end{figure*}

In this work, we study the modified magnetization dynamics of a FI adjacent to graphene. 
Figure \ref{fig_system} (a) shows a schematic of the system. Microwaves are irradiated and the precession of localized spins is induced.
Figure \ref{fig_system} (b) and (c) shows the electronic structure of graphene on the FI under a perpendicular magnetic field. The spin degeneracy is lifted by the exchange coupling at the interface and spin-split Landau levels are formed. The densities of states for spin-up and spin-down are shown in the right panel; Landau level broadening is included. We find that the modulation of Gilbert damping is proportional to the product of the densities of states for spin-up and spin-down, so that the FMR measurements may be used as a probe of the spin-resolved densities of states. Owing to the peak structure of the density of states, the modulation of Gilbert damping exhibits peak structure and an oscillation as a function of Fermi level and magnetic field, which reflects the Landau level structure. One may determine the exchange coupling constant by analyzing the period of the oscillation.

{\it Model Hamiltonian}.---The total Hamiltonian $H(t)$ consists of three terms,
\begin{align}
    H(t)=\hfi(t)+H_{\rm Gr}+\hex.
    \label{eq_Hamiltonian}
\end{align}
The first term $\hfi(t)$ describes the bulk FI 
\begin{align}
\hfi(t)=
    &\sum_\bk\ho_\bk \bc\ba
    -\hac^+(t)b^\dagger_{\bk=\bm{0}}-\hac^-(t)b_{\bk=\bm{0}},
\end{align}
where $\bc$ and $\ba$ denote the creation and annihilation operators of magnons with momentum $\bk$.
We assume a parabolic dispersion $\hbar\omega_\bk=Dk^2-\hbar\gamma B$, with $\gamma(<0)$ the electron gyromagnetic ratio.
The coupling between the microwave and magnons is given by
\begin{align}
    \hac^\pm(t)=\frac{\hbar\gamma\hac}{2}\sqrt{2SN}e^{\mp i\Omega t},
\end{align}
where $\hac$ and $\Omega$ are the amplitude and frequency of the microwave radiation, respectively, and $S$ is the magnitude of the localized spin in the FI.
The above Hamiltonian is derived from a ferromagnetic Heisenberg model using the Holstein-Primakoff transformation and the spin-wave approximation ($S^z_\bk=S-b^\dagger_\bk b_\bk$, $S_{\bk}^+=\sqrt{2S}\hspace{0.5mm}\ba$, $S_{-\bk}^{-}=\sqrt{2S}\hspace{0.5mm}\bc$, where $\bm{S}_\bk$ is the Fourier transform of the localized spin in the FI).

The second term $H_{\rm Gr}$ describes the electronic states around the $K$ point in graphene under a perpendicular magnetic field,
\begin{align}
    H_{\rm Gr}=\sum_{nXs} \e_{n} c_{nXs}^\dagger c_{nXs},
\end{align}
where $c_{nXs}^\dagger$ and $c_{nXs}$ denote the creation and annihilation operators of electrons with Landau level index $n=0,\pm1,\pm2,\cdots$, guiding center $X$, and spin up $s=+$ and spin down $s=-$. The eigenenergy is given by
\begin{align}
    \e_{n}={\rm sgn}(n)\sqrt{2e\hbar v^2}\sqrt{|n|B},
\end{align}
where $v$ is the velocity and the sign function is defined as
\begin{align}
    {\rm sgn}(n):=
        \begin{cases}
            1  &(n>0) \\
            0  &(n=0) \\
            -1 &(n<0)
        \end{cases}.
\end{align}
In the following, we neglect the Zeeman coupling between the electron spin and the magnetic field because it is much smaller than the Landau-level separation and the exchange coupling introduced below.
In graphene, there are two inequivalent valleys labelled $K$ and $K^\p$. In this paper, we assume that the intervalley scattering is negligible. This assumption is valid for an atomically flat interface, which is reasonable given the recent experimental setups \cite{wangProximityInducedFerromagnetismGraphene2015a,mendesSpinCurrentChargeCurrentConversion2015,mendesDirectDetectionInduced2019}. Consequently, the valley degree of freedom just doubles the modulation of the Gilbert damping. 

The third term $\hex$ is the exchange coupling at the interface consisting of two terms
\begin{align}
    \hex=H_{\rm Z}+H_{\rm T},
\end{align}
where $H_{\rm Z}$ denotes the out-of-plane component of the exchange coupling and leads to the spin splitting in graphene,
\begin{align}
    &H_{\rm Z}
    =-JS\sum_{nX}
    \left(
        c_{nX+}^\dagger
        c_{nX+}
        -
        c_{nX-}^\dagger
        c_{nX-}
    \right),
\end{align}
with $J$ the exchange coupling constant.
The $z$-component of the localized spin is approximated as $\la S^z_\bk\ra\approx S$.
The out-of-plane component $H_{\rm Z}$ is modeled as a uniform Zeeman-like coupling, although in general, $H_{\rm Z}$ contains the effect of surface roughness, which gives off-diagonal terms.
The Hamiltonian $H_{\rm T}$ denotes the in-plane component of the exchange coupling and describes spin transfer between the FI and graphene,
\begin{align}
    &H_{\rm T}
    =-\sum_{nX}\sum_{n^\p X^\p}\sum_\bk
        \left(
            J_{nX,n^\p X^\p,\bk}s^+_{nX+,n^\p X^\p-}S^-_\bk+{\rm h.c.}
        \right),
\end{align}
where $J_{nX,n^\p X^\p,\bk}$ is the matrix element for the spin transfer processes and $s^+_{nX+,n^\p X^\p-}$ is the spin-flip operator for the electron spin in graphene.

{\it Modulation of Gilbert Damping}.---To discuss the Gilbert damping, we calculated the time-dependent statistical average of the localized spin under the microwave irradiation. The first-order perturbation calculation gives the deviation from the thermal average,
\begin{align}
    \d\la S^+_{\bk=\bm{0}}(t)\ra=
        -\hac^+(t)G^R_{\bk=\bm{0}}(\Omega).
\end{align}
The retarded Green's function is written as
\begin{align}
    &G_\bk^R(\omega)
        =\frac
            {2S/\hbar}
            {
                \omega
                -\omega_\bk
                +i\ag\omega
                -(2S/\hbar)\Sigma^R_\bk(\omega)
            },
\end{align}
where we have introduced the phenomenological dimensionless damping parameter $\ag$, called the Gilbert damping constant, which originates from the magnon-phonon and magnon-magnon coupling, etc \cite{kasuyaRelaxationMechanismsFerromagnetic1961,cherepanovSagaYIGSpectra1993,jinTemperatureDependenceSpinwave2019}.
In this paper, we focus on the modulation of the Gilbert damping stemming from the spin transfer processes at the interface.
The self-energy from the spin transfer processes at the interface within second-order perturbation is given by
\begin{align}
    \Sigma^R_\bk(\omega)
        &=\sum_{nX}\sum_{n^\p X^\p}|J_{nX,n^\p X^\p,\bk}|^2
            \chi_{n+,n^\p-}^R(\omega).
\end{align}
The spin susceptibility is given by
\begin{align}
    \chi_{n+,n^\p-}^R(\omega)=\frac{f_{n+}-f_{n^\p-}}{\e_{n+}-\e_{n^\p-}+\ho+i0},
\end{align}
where $f_{ns}=1/\left(e^{(\e_{ns}-\mu)/k_{\rm B}T}+1\right)$ is the Fermi distribution function and $\e_{ns}=\e_n-JSs$ is the spin-split Landau level.
From the self-energy expression, one sees that the modulation of the Gilbert damping reflects the property of the spin susceptibility of graphene.
The modulation of the Gilbert damping under the microwave irradiation is given by \cite{simanekGilbertDampingMagnetic2003,ohnumaEnhancedDcSpin2014,tataraConsistentMicroscopicAnalysis2017,inoueSpinPumpingSuperconductors2017,matsuoSpinCurrentNoise2018,katoMicroscopicTheorySpin2019}
\begin{align}
    \d\ag^K=
        -\frac{2S{\rm Im}\Sigma_{\bk=\bm{0}}^R(\omega)}{\hbar\omega},
\end{align}
where the superscript $K$ signifies the contribution from the $K$ valley.

To further the calculation, we assume that the matrix element $J_{nX,n^\p X^\p,\bk=\bm{0}}$ is approximated by a constant $J_0$, including detail properties of the interface, that is, $J_{nX,n^\p X^\p,\bk=\bm{0}}\approx J_0$.
With this assumption, the self-energy becomes
\begin{align}
    {\rm Im}\Sigma_{\bk=\bm{0}}^R(\omega)
    &=-|J_0|^2\pi\ho
        \int d\e\left(-\frac{\partial f(\e)}{\partial\e}\right)
        D_+(\e)D_-(\e),
\end{align}
where $D_s(\e)$ is the density of states for spin $s=\pm$
\begin{align}
    D_{s}(\e)
        =
        \frac{A}{2\pi\lb^2}
        \sum_n
        \frac{1}{\pi}
        \frac{\Gamma}{(\e-\e_{ns})^2+\Gamma^2},
\end{align}
with magnetic length $\lb=\sqrt{\hbar/(eB)}$ and area of the interface $A$.
Here, we have introduced a constant $\Gamma$ describing level broadening arising from surface roughness and impurity scattering. This is the simplest approximation to include the disorder effect.
The density of states shows peaks at the Landau level, which is prominent when its separation exceeds the level broadening.
Finally, the modulation of the Gilbert damping constant $\d\ag$ is derived as
\begin{align}
    &\d\ag=
        2\pi g_vS|J_0|^2     
        \int d\e\left(-\frac{\partial f(\e)}{\partial\e}\right)
        D_{+}(\e)D_{-}(\e),
\end{align}
where $g_v=2$ denotes the valley degree of freedom.
From this expression, one sees that the modulation of the Gilbert damping is proportional to the product of the densities of states for spin-up and spin-down.
Therefore, combined with the density of states measurement, for example, a capacitance measurement \cite{ponomarenkoDensityStatesZero2010}, the FMR measurement is used to detect the spin-resolved densities of states.

\begin{figure}
\begin{center}
\includegraphics[width=1\hsize]{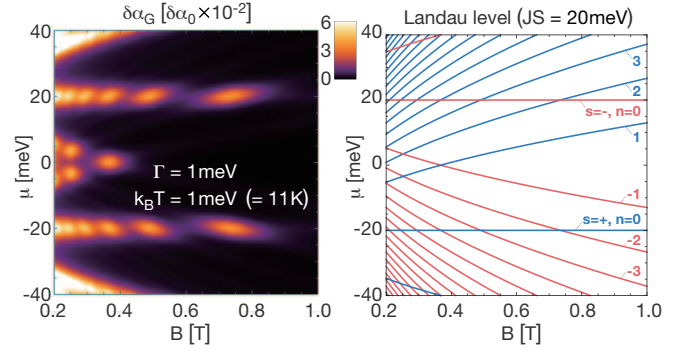}
\end{center}
\caption{
(Color online) Modulation of the Gilbert damping constant $\d\ag$ and spin-split Landau levels as a function of the Fermi level $\mu$ and the magnetic field $B$.
The spin splitting $JS$ is set to $20{\rm meV}$.
In the left panel, $\d\ag$ has peaks at the crossing points of spin-up and spin-down Landau levels.
In the right panel, the blue and red curves identify the spin-up and spin-down Landau levels, respectively.
}
\label{fig_DensityPlot}
\end{figure}

Figure \ref{fig_DensityPlot} shows the spin-split Landau levels and the modulation of the Gilbert damping $\d\ag$ as a function of the Fermi level $\mu$ and the magnetic field $B$.
We use $\d\a_0$ as a unit of $\d\ag$
\begin{align}
    &\d\a_0=2\pi g_vS|J_0|^2\left(\frac{A}{2\pi\lb^2}\frac{1}{\rm meV}\right)^2.
\end{align}
We note that $\d\a_0(\propto B^2)$ depends on the magnetic field.
Both the level broadening $\Gamma$ and the thermal broadening $k_{\rm B}T$ are set to $1{\rm meV}$, and $JS$ is set to $20{\rm meV}$
\cite{yangProximityEffectsInduced2013,wangProximityInducedFerromagnetismGraphene2015a,hallalTailoringMagneticInsulator2017}.
$\d\ag$ reflects the Landau level structure and has peaks at crossing points of spin-up and spin-down Landau levels.
The peak positions are determined by solving $\e_{n+}=\e_{n^\p-}$ and the inverse of the magnetic field at the peaks is given by
\begin{align}
    \frac{1}{B}=\frac{2e\hbar v^2}{(2JS)^2}\left(\sqrt{|n|}-\sqrt{|n^\p|}\right)^2.
    \label{eq_Peak_B}
\end{align}
The peak structure becomes prominent when the Landau level separation exceeds both level and thermal broadening.

\begin{figure}
\begin{center}
\includegraphics[width=1\hsize]{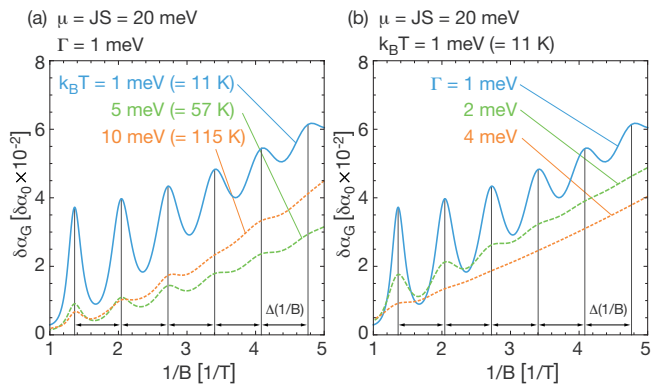}
\end{center}
\caption{
(Color online) Quantum oscillation of the modulation of the Gilbert damping constant $\d\ag$ as a function of the inverse of the magnetic field $1/B$.
The Fermi level $\mu$ and the magnitude of the spin splitting $JS$ are set to $20{\rm meV}$.
(a) $\Gamma=1{\rm meV}$ and $\d\ag$ is plotted at several temperatures.
(b) $k_{\rm B}T=1{\rm meV}$ and $\d\ag$ is plotted for several $\Gamma$'s. The period of the oscillation $\Delta(1/B)$ is indicated by double-headed arrows.
}
\label{fig_Oscillation}
\end{figure}

Figure \ref{fig_Oscillation} shows the modulation of the Gilbert damping $\d\ag$ as a function of the inverse of the magnetic field $1/B$ with the Fermi level set to $\mu=20{\rm meV}$, where the spin-down zeroth Landau level resides.
$\d\ag$ shows peak structure and a periodic oscillatory behavior.
The period of the oscillation $\Delta(1/B)$ is derived from Eq.\ (\ref{eq_Peak_B}) and is written as
\begin{align}
    \Delta\left(\frac{1}{B}\right)=\frac{2e\hbar v^2}{(2JS)^2}.
\end{align}
The above relation means that the magnitude of the spin splitting $JS$ is detectable by measuring the period of the oscillation $\Delta(1/B)$.
For the peak structure to be clear, both level and thermal broadening must to be sufficiently smaller than the Landau level separation; otherwise, the peak structure smears out.

{\it Discussion}.---To observe the oscillation of Gilbert damping, at least two conditions must be satisfied.
First, the well-separated landau levels have to be realized in the magnetic field where the FMR measurements is feasible.
Second, the FMR modulation caused by the adjacent graphene have to be detectable.
The graphene Landau levels are observed in recent experiments at $2{\rm T}$ \cite{banszerusUltrahighmobilityGrapheneDevices2015}, and recent broadband ferromagnetic resonance spectrometer enables the generation of microwaves with frequencies $\leq40{\rm GHz}$ and FMR measurements in a magnetic field $\leq2{\rm T}$ \cite{ohshimaSpinInjectionSilicon2017}.
The modulation of the FMR linewidth in Permalloy/Graphene \cite{tangDynamicallyGeneratedPure2013,indoleseWidebandOnChipExcitation2018}, yttrium iron garnet/Graphene \cite{mendesSpinCurrentChargeCurrentConversion2015,mendesDirectDetectionInduced2019} have been reported by several experimental groups, although all of them were performed at room temperature.
Therefore, the above two conditions are experimentally feasible and our theoretical predictions can be tested in an appropriate experimental setup.

{\it Conclusion}.---We have studied the modulation of the Gilbert damping $\d\ag$ in a ferromagnetic insulator on which graphene is placed. The exchange coupling at the interface and the perpendicular magnetic field lead to the spin-split Landau levels in graphene. We showed that $\d\ag$ is proportional to the product of the densities of states for spin-up and spin-down electrons. Therefore, the spin-resolved densities of states can be detected by measuring $\d\ag$ and the total density of states.
When the Fermi level is fixed at a Landau level, $\d\ag$ oscillates as a function of the inverse of the magnetic field. The period of the oscillation provides information on the magnitude of the spin splitting.
Our main message is that the FMR measurement is a probe of spin-resolved electronic structure.
In addition to spin current generation, one may use the FMR measurements to detect the electronic structure of adjacent materials.

{\bf Acnowledgement}
We thank J. Fujimoto, T. Kato, R. Ohshima, and M. Shiraishi for helpful discussions.
This work is partially supported by the Priority Program of Chinese Academy of Sciences, Grant No. XDB28000000.

\bibliographystyle{apsrev4-1}
\bibliography{17657}

\begin{thebibliography}{37}%
\makeatletter
\providecommand \@ifxundefined [1]{%
 \@ifx{#1\undefined}
}%
\providecommand \@ifnum [1]{%
 \ifnum #1\expandafter \@firstoftwo
 \else \expandafter \@secondoftwo
 \fi
}%
\providecommand \@ifx [1]{%
 \ifx #1\expandafter \@firstoftwo
 \else \expandafter \@secondoftwo
 \fi
}%
\providecommand \natexlab [1]{#1}%
\providecommand \enquote  [1]{``#1''}%
\providecommand \bibnamefont  [1]{#1}%
\providecommand \bibfnamefont [1]{#1}%
\providecommand \citenamefont [1]{#1}%
\providecommand \href@noop [0]{\@secondoftwo}%
\providecommand \href [0]{\begingroup \@sanitize@url \@href}%
\providecommand \@href[1]{\@@startlink{#1}\@@href}%
\providecommand \@@href[1]{\endgroup#1\@@endlink}%
\providecommand \@sanitize@url [0]{\catcode `\\12\catcode `\$12\catcode
  `\&12\catcode `\#12\catcode `\^12\catcode `\_12\catcode `\%12\relax}%
\providecommand \@@startlink[1]{}%
\providecommand \@@endlink[0]{}%
\providecommand \url  [0]{\begingroup\@sanitize@url \@url }%
\providecommand \@url [1]{\endgroup\@href {#1}{\urlprefix }}%
\providecommand \urlprefix  [0]{URL }%
\providecommand \Eprint [0]{\href }%
\providecommand \doibase [0]{http://dx.doi.org/}%
\providecommand \selectlanguage [0]{\@gobble}%
\providecommand \bibinfo  [0]{\@secondoftwo}%
\providecommand \bibfield  [0]{\@secondoftwo}%
\providecommand \translation [1]{[#1]}%
\providecommand \BibitemOpen [0]{}%
\providecommand \bibitemStop [0]{}%
\providecommand \bibitemNoStop [0]{.\EOS\space}%
\providecommand \EOS [0]{\spacefactor3000\relax}%
\providecommand \BibitemShut  [1]{\csname bibitem#1\endcsname}%
\let\auto@bib@innerbib\@empty
\bibitem [{\citenamefont {Han}\ \emph {et~al.}(2014)\citenamefont {Han},
  \citenamefont {Kawakami}, \citenamefont {Gmitra},\ and\ \citenamefont
  {Fabian}}]{hanGrapheneSpintronics2014}%
  \BibitemOpen
  \bibfield  {author} {\bibinfo {author} {\bibfnamefont {W.}~\bibnamefont
  {Han}}, \bibinfo {author} {\bibfnamefont {R.~K.}\ \bibnamefont {Kawakami}},
  \bibinfo {author} {\bibfnamefont {M.}~\bibnamefont {Gmitra}}, \ and\ \bibinfo
  {author} {\bibfnamefont {J.}~\bibnamefont {Fabian}},\ }\href {\doibase
  10.1038/nnano.2014.214} {\bibfield  {journal} {\bibinfo  {journal} {Nature
  Nanotechnology}\ }\textbf {\bibinfo {volume} {9}},\ \bibinfo {pages} {794}
  (\bibinfo {year} {2014})}\BibitemShut {NoStop}%
\bibitem [{\citenamefont {Yang}\ \emph {et~al.}(2013)\citenamefont {Yang},
  \citenamefont {Hallal}, \citenamefont {Terrade}, \citenamefont {Waintal},
  \citenamefont {Roche},\ and\ \citenamefont
  {Chshiev}}]{yangProximityEffectsInduced2013}%
  \BibitemOpen
  \bibfield  {author} {\bibinfo {author} {\bibfnamefont {H.~X.}\ \bibnamefont
  {Yang}}, \bibinfo {author} {\bibfnamefont {A.}~\bibnamefont {Hallal}},
  \bibinfo {author} {\bibfnamefont {D.}~\bibnamefont {Terrade}}, \bibinfo
  {author} {\bibfnamefont {X.}~\bibnamefont {Waintal}}, \bibinfo {author}
  {\bibfnamefont {S.}~\bibnamefont {Roche}}, \ and\ \bibinfo {author}
  {\bibfnamefont {M.}~\bibnamefont {Chshiev}},\ }\href {\doibase
  10.1103/PhysRevLett.110.046603} {\bibfield  {journal} {\bibinfo  {journal}
  {Physical Review Letters}\ }\textbf {\bibinfo {volume} {110}},\ \bibinfo
  {pages} {046603} (\bibinfo {year} {2013})}\BibitemShut {NoStop}%
\bibitem [{\citenamefont {Hallal}\ \emph {et~al.}(2017)\citenamefont {Hallal},
  \citenamefont {Ibrahim}, \citenamefont {Yang}, \citenamefont {Roche},\ and\
  \citenamefont {Chshiev}}]{hallalTailoringMagneticInsulator2017}%
  \BibitemOpen
  \bibfield  {author} {\bibinfo {author} {\bibfnamefont {A.}~\bibnamefont
  {Hallal}}, \bibinfo {author} {\bibfnamefont {F.}~\bibnamefont {Ibrahim}},
  \bibinfo {author} {\bibfnamefont {H.}~\bibnamefont {Yang}}, \bibinfo {author}
  {\bibfnamefont {S.}~\bibnamefont {Roche}}, \ and\ \bibinfo {author}
  {\bibfnamefont {M.}~\bibnamefont {Chshiev}},\ }\href {\doibase
  10.1088/2053-1583/aa6663} {\bibfield  {journal} {\bibinfo  {journal} {2D
  Materials}\ }\textbf {\bibinfo {volume} {4}},\ \bibinfo {pages} {025074}
  (\bibinfo {year} {2017})}\BibitemShut {NoStop}%
\bibitem [{\citenamefont {Wang}\ \emph {et~al.}(2015)\citenamefont {Wang},
  \citenamefont {Tang}, \citenamefont {Sachs}, \citenamefont {Barlas},\ and\
  \citenamefont {Shi}}]{wangProximityInducedFerromagnetismGraphene2015a}%
  \BibitemOpen
  \bibfield  {author} {\bibinfo {author} {\bibfnamefont {Z.}~\bibnamefont
  {Wang}}, \bibinfo {author} {\bibfnamefont {C.}~\bibnamefont {Tang}}, \bibinfo
  {author} {\bibfnamefont {R.}~\bibnamefont {Sachs}}, \bibinfo {author}
  {\bibfnamefont {Y.}~\bibnamefont {Barlas}}, \ and\ \bibinfo {author}
  {\bibfnamefont {J.}~\bibnamefont {Shi}},\ }\href {\doibase
  10.1103/PhysRevLett.114.016603} {\bibfield  {journal} {\bibinfo  {journal}
  {Physical Review Letters}\ }\textbf {\bibinfo {volume} {114}},\ \bibinfo
  {pages} {016603} (\bibinfo {year} {2015})}\BibitemShut {NoStop}%
\bibitem [{\citenamefont {Averyanov}\ \emph {et~al.}(2018)\citenamefont
  {Averyanov}, \citenamefont {Sokolov}, \citenamefont {Tokmachev},
  \citenamefont {Parfenov}, \citenamefont {Karateev}, \citenamefont
  {Taldenkov},\ and\ \citenamefont
  {Storchak}}]{averyanovHighTemperatureMagnetismGraphene2018}%
  \BibitemOpen
  \bibfield  {author} {\bibinfo {author} {\bibfnamefont {D.~V.}\ \bibnamefont
  {Averyanov}}, \bibinfo {author} {\bibfnamefont {I.~S.}\ \bibnamefont
  {Sokolov}}, \bibinfo {author} {\bibfnamefont {A.~M.}\ \bibnamefont
  {Tokmachev}}, \bibinfo {author} {\bibfnamefont {O.~E.}\ \bibnamefont
  {Parfenov}}, \bibinfo {author} {\bibfnamefont {I.~A.}\ \bibnamefont
  {Karateev}}, \bibinfo {author} {\bibfnamefont {A.~N.}\ \bibnamefont
  {Taldenkov}}, \ and\ \bibinfo {author} {\bibfnamefont {V.~G.}\ \bibnamefont
  {Storchak}},\ }\href {\doibase 10.1021/acsami.8b04289} {\bibfield  {journal}
  {\bibinfo  {journal} {ACS Applied Materials \& Interfaces}\ }\textbf
  {\bibinfo {volume} {10}},\ \bibinfo {pages} {20767} (\bibinfo {year}
  {2018})}\BibitemShut {NoStop}%
\bibitem [{\citenamefont {Leutenantsmeyer}\ \emph {et~al.}(2016)\citenamefont
  {Leutenantsmeyer}, \citenamefont {Kaverzin}, \citenamefont {Wojtaszek},\ and\
  \citenamefont {van Wees}}]{leutenantsmeyerProximityInducedRoom2016}%
  \BibitemOpen
  \bibfield  {author} {\bibinfo {author} {\bibfnamefont {J.~C.}\ \bibnamefont
  {Leutenantsmeyer}}, \bibinfo {author} {\bibfnamefont {A.~A.}\ \bibnamefont
  {Kaverzin}}, \bibinfo {author} {\bibfnamefont {M.}~\bibnamefont {Wojtaszek}},
  \ and\ \bibinfo {author} {\bibfnamefont {B.~J.}\ \bibnamefont {van Wees}},\
  }\href {\doibase 10.1088/2053-1583/4/1/014001} {\bibfield  {journal}
  {\bibinfo  {journal} {2D Materials}\ }\textbf {\bibinfo {volume} {4}},\
  \bibinfo {pages} {014001} (\bibinfo {year} {2016})}\BibitemShut {NoStop}%
\bibitem [{\citenamefont {Wei}\ \emph {et~al.}(2016)\citenamefont {Wei},
  \citenamefont {Lee}, \citenamefont {Lemaitre}, \citenamefont {Pinel},
  \citenamefont {Cutaia}, \citenamefont {Cha}, \citenamefont {Katmis},
  \citenamefont {Zhu}, \citenamefont {Heiman}, \citenamefont {Hone},
  \citenamefont {Moodera},\ and\ \citenamefont
  {Chen}}]{weiStrongInterfacialExchange2016a}%
  \BibitemOpen
  \bibfield  {author} {\bibinfo {author} {\bibfnamefont {P.}~\bibnamefont
  {Wei}}, \bibinfo {author} {\bibfnamefont {S.}~\bibnamefont {Lee}}, \bibinfo
  {author} {\bibfnamefont {F.}~\bibnamefont {Lemaitre}}, \bibinfo {author}
  {\bibfnamefont {L.}~\bibnamefont {Pinel}}, \bibinfo {author} {\bibfnamefont
  {D.}~\bibnamefont {Cutaia}}, \bibinfo {author} {\bibfnamefont
  {W.}~\bibnamefont {Cha}}, \bibinfo {author} {\bibfnamefont {F.}~\bibnamefont
  {Katmis}}, \bibinfo {author} {\bibfnamefont {Y.}~\bibnamefont {Zhu}},
  \bibinfo {author} {\bibfnamefont {D.}~\bibnamefont {Heiman}}, \bibinfo
  {author} {\bibfnamefont {J.}~\bibnamefont {Hone}}, \bibinfo {author}
  {\bibfnamefont {J.~S.}\ \bibnamefont {Moodera}}, \ and\ \bibinfo {author}
  {\bibfnamefont {C.-T.}\ \bibnamefont {Chen}},\ }\href {\doibase
  10.1038/nmat4603} {\bibfield  {journal} {\bibinfo  {journal} {Nature
  Materials}\ }\textbf {\bibinfo {volume} {15}},\ \bibinfo {pages} {711}
  (\bibinfo {year} {2016})}\BibitemShut {NoStop}%
\bibitem [{\citenamefont {Mizukami}\ \emph
  {et~al.}(2001{\natexlab{a}})\citenamefont {Mizukami}, \citenamefont {Ando},\
  and\ \citenamefont {Miyazaki}}]{mizukamiStudyFerromagneticResonance2001}%
  \BibitemOpen
  \bibfield  {author} {\bibinfo {author} {\bibfnamefont {S.}~\bibnamefont
  {Mizukami}}, \bibinfo {author} {\bibfnamefont {Y.}~\bibnamefont {Ando}}, \
  and\ \bibinfo {author} {\bibfnamefont {T.}~\bibnamefont {Miyazaki}},\ }\href
  {\doibase 10.1143/JJAP.40.580} {\bibfield  {journal} {\bibinfo  {journal}
  {Japanese Journal of Applied Physics}\ }\textbf {\bibinfo {volume} {40}},\
  \bibinfo {pages} {580} (\bibinfo {year} {2001}{\natexlab{a}})}\BibitemShut
  {NoStop}%
\bibitem [{\citenamefont {Mizukami}\ \emph
  {et~al.}(2001{\natexlab{b}})\citenamefont {Mizukami}, \citenamefont {Ando},\
  and\ \citenamefont {Miyazaki}}]{mizukamiFerromagneticResonanceLinewidth2001}%
  \BibitemOpen
  \bibfield  {author} {\bibinfo {author} {\bibfnamefont {S.}~\bibnamefont
  {Mizukami}}, \bibinfo {author} {\bibfnamefont {Y.}~\bibnamefont {Ando}}, \
  and\ \bibinfo {author} {\bibfnamefont {T.}~\bibnamefont {Miyazaki}},\
  }\href@noop {} {\bibfield  {journal} {\bibinfo  {journal} {Journal of
  Magnetism and Magnetic Materials}\ }\textbf {\bibinfo {volume} {226}},\
  \bibinfo {pages} {1640} (\bibinfo {year} {2001}{\natexlab{b}})}\BibitemShut
  {NoStop}%
\bibitem [{\citenamefont {Mizukami}\ \emph {et~al.}(2002)\citenamefont
  {Mizukami}, \citenamefont {Ando},\ and\ \citenamefont
  {Miyazaki}}]{mizukamiEffectSpinDiffusion2002}%
  \BibitemOpen
  \bibfield  {author} {\bibinfo {author} {\bibfnamefont {S.}~\bibnamefont
  {Mizukami}}, \bibinfo {author} {\bibfnamefont {Y.}~\bibnamefont {Ando}}, \
  and\ \bibinfo {author} {\bibfnamefont {T.}~\bibnamefont {Miyazaki}},\ }\href
  {\doibase 10.1103/PhysRevB.66.104413} {\bibfield  {journal} {\bibinfo
  {journal} {Physical Review B}\ }\textbf {\bibinfo {volume} {66}},\ \bibinfo
  {pages} {104413} (\bibinfo {year} {2002})}\BibitemShut {NoStop}%
\bibitem [{\citenamefont {Rahimi}\ and\ \citenamefont
  {Moghaddam}(2015)}]{rahimiElectricallyControllableSpin2015}%
  \BibitemOpen
  \bibfield  {author} {\bibinfo {author} {\bibfnamefont {M.~A.}\ \bibnamefont
  {Rahimi}}\ and\ \bibinfo {author} {\bibfnamefont {A.~G.}\ \bibnamefont
  {Moghaddam}},\ }\href {\doibase 10.1088/0022-3727/48/29/295004} {\bibfield
  {journal} {\bibinfo  {journal} {Journal of Physics D: Applied Physics}\
  }\textbf {\bibinfo {volume} {48}},\ \bibinfo {pages} {295004} (\bibinfo
  {year} {2015})}\BibitemShut {NoStop}%
\bibitem [{\citenamefont {Inoue}\ \emph {et~al.}(2016)\citenamefont {Inoue},
  \citenamefont {Bauer},\ and\ \citenamefont
  {Nomura}}]{inoueSpinPumpingTwodimensional2016}%
  \BibitemOpen
  \bibfield  {author} {\bibinfo {author} {\bibfnamefont {T.}~\bibnamefont
  {Inoue}}, \bibinfo {author} {\bibfnamefont {G.~E.~W.}\ \bibnamefont {Bauer}},
  \ and\ \bibinfo {author} {\bibfnamefont {K.}~\bibnamefont {Nomura}},\ }\href
  {\doibase 10.1103/PhysRevB.94.205428} {\bibfield  {journal} {\bibinfo
  {journal} {Physical Review B}\ }\textbf {\bibinfo {volume} {94}},\ \bibinfo
  {pages} {205428} (\bibinfo {year} {2016})}\BibitemShut {NoStop}%
\bibitem [{\citenamefont {Patra}\ \emph {et~al.}(2012)\citenamefont {Patra},
  \citenamefont {Singh}, \citenamefont {Barin}, \citenamefont {Lee},
  \citenamefont {Ahn}, \citenamefont {{del Barco}}, \citenamefont {Mucciolo},\
  and\ \citenamefont {{\"O}zyilmaz}}]{patraDynamicSpinInjection2012}%
  \BibitemOpen
  \bibfield  {author} {\bibinfo {author} {\bibfnamefont {A.~K.}\ \bibnamefont
  {Patra}}, \bibinfo {author} {\bibfnamefont {S.}~\bibnamefont {Singh}},
  \bibinfo {author} {\bibfnamefont {B.}~\bibnamefont {Barin}}, \bibinfo
  {author} {\bibfnamefont {Y.}~\bibnamefont {Lee}}, \bibinfo {author}
  {\bibfnamefont {J.-H.}\ \bibnamefont {Ahn}}, \bibinfo {author} {\bibfnamefont
  {E.}~\bibnamefont {{del Barco}}}, \bibinfo {author} {\bibfnamefont {E.~R.}\
  \bibnamefont {Mucciolo}}, \ and\ \bibinfo {author} {\bibfnamefont
  {B.}~\bibnamefont {{\"O}zyilmaz}},\ }\href {\doibase 10.1063/1.4761932}
  {\bibfield  {journal} {\bibinfo  {journal} {Applied Physics Letters}\
  }\textbf {\bibinfo {volume} {101}},\ \bibinfo {pages} {162407} (\bibinfo
  {year} {2012})}\BibitemShut {NoStop}%
\bibitem [{\citenamefont {Tang}\ \emph {et~al.}(2013)\citenamefont {Tang},
  \citenamefont {Shikoh}, \citenamefont {Ago}, \citenamefont {Kawahara},
  \citenamefont {Ando}, \citenamefont {Shinjo},\ and\ \citenamefont
  {Shiraishi}}]{tangDynamicallyGeneratedPure2013}%
  \BibitemOpen
  \bibfield  {author} {\bibinfo {author} {\bibfnamefont {Z.}~\bibnamefont
  {Tang}}, \bibinfo {author} {\bibfnamefont {E.}~\bibnamefont {Shikoh}},
  \bibinfo {author} {\bibfnamefont {H.}~\bibnamefont {Ago}}, \bibinfo {author}
  {\bibfnamefont {K.}~\bibnamefont {Kawahara}}, \bibinfo {author}
  {\bibfnamefont {Y.}~\bibnamefont {Ando}}, \bibinfo {author} {\bibfnamefont
  {T.}~\bibnamefont {Shinjo}}, \ and\ \bibinfo {author} {\bibfnamefont
  {M.}~\bibnamefont {Shiraishi}},\ }\href {\doibase 10.1103/PhysRevB.87.140401}
  {\bibfield  {journal} {\bibinfo  {journal} {Physical Review B}\ }\textbf
  {\bibinfo {volume} {87}},\ \bibinfo {pages} {140401} (\bibinfo {year}
  {2013})}\BibitemShut {NoStop}%
\bibitem [{\citenamefont {Dushenko}\ \emph {et~al.}(2016)\citenamefont
  {Dushenko}, \citenamefont {Ago}, \citenamefont {Kawahara}, \citenamefont
  {Tsuda}, \citenamefont {Kuwabata}, \citenamefont {Takenobu}, \citenamefont
  {Shinjo}, \citenamefont {Ando},\ and\ \citenamefont
  {Shiraishi}}]{dushenkoGateTunableSpinChargeConversion2016}%
  \BibitemOpen
  \bibfield  {author} {\bibinfo {author} {\bibfnamefont {S.}~\bibnamefont
  {Dushenko}}, \bibinfo {author} {\bibfnamefont {H.}~\bibnamefont {Ago}},
  \bibinfo {author} {\bibfnamefont {K.}~\bibnamefont {Kawahara}}, \bibinfo
  {author} {\bibfnamefont {T.}~\bibnamefont {Tsuda}}, \bibinfo {author}
  {\bibfnamefont {S.}~\bibnamefont {Kuwabata}}, \bibinfo {author}
  {\bibfnamefont {T.}~\bibnamefont {Takenobu}}, \bibinfo {author}
  {\bibfnamefont {T.}~\bibnamefont {Shinjo}}, \bibinfo {author} {\bibfnamefont
  {Y.}~\bibnamefont {Ando}}, \ and\ \bibinfo {author} {\bibfnamefont
  {M.}~\bibnamefont {Shiraishi}},\ }\href {\doibase
  10.1103/PhysRevLett.116.166102} {\bibfield  {journal} {\bibinfo  {journal}
  {Physical Review Letters}\ }\textbf {\bibinfo {volume} {116}},\ \bibinfo
  {pages} {166102} (\bibinfo {year} {2016})}\BibitemShut {NoStop}%
\bibitem [{\citenamefont {Indolese}\ \emph {et~al.}(2018)\citenamefont
  {Indolese}, \citenamefont {Zihlmann}, \citenamefont {Makk}, \citenamefont
  {J{\"u}nger}, \citenamefont {Thodkar},\ and\ \citenamefont
  {Sch{\"o}nenberger}}]{indoleseWidebandOnChipExcitation2018}%
  \BibitemOpen
  \bibfield  {author} {\bibinfo {author} {\bibfnamefont {D.}~\bibnamefont
  {Indolese}}, \bibinfo {author} {\bibfnamefont {S.}~\bibnamefont {Zihlmann}},
  \bibinfo {author} {\bibfnamefont {P.}~\bibnamefont {Makk}}, \bibinfo {author}
  {\bibfnamefont {C.}~\bibnamefont {J{\"u}nger}}, \bibinfo {author}
  {\bibfnamefont {K.}~\bibnamefont {Thodkar}}, \ and\ \bibinfo {author}
  {\bibfnamefont {C.}~\bibnamefont {Sch{\"o}nenberger}},\ }\href {\doibase
  10.1103/PhysRevApplied.10.044053} {\bibfield  {journal} {\bibinfo  {journal}
  {Physical Review Applied}\ }\textbf {\bibinfo {volume} {10}},\ \bibinfo
  {pages} {044053} (\bibinfo {year} {2018})}\BibitemShut {NoStop}%
\bibitem [{\citenamefont {Mendes}\ \emph {et~al.}(2015)\citenamefont {Mendes},
  \citenamefont {Alves~Santos}, \citenamefont {Meireles}, \citenamefont
  {Lacerda}, \citenamefont {{Vilela-Le{\~a}o}}, \citenamefont {Machado},
  \citenamefont {{Rodr{\'i}guez-Su{\'a}rez}}, \citenamefont {Azevedo},\ and\
  \citenamefont {Rezende}}]{mendesSpinCurrentChargeCurrentConversion2015}%
  \BibitemOpen
  \bibfield  {author} {\bibinfo {author} {\bibfnamefont {J.~B.~S.}\
  \bibnamefont {Mendes}}, \bibinfo {author} {\bibfnamefont {O.}~\bibnamefont
  {Alves~Santos}}, \bibinfo {author} {\bibfnamefont {L.~M.}\ \bibnamefont
  {Meireles}}, \bibinfo {author} {\bibfnamefont {R.~G.}\ \bibnamefont
  {Lacerda}}, \bibinfo {author} {\bibfnamefont {L.~H.}\ \bibnamefont
  {{Vilela-Le{\~a}o}}}, \bibinfo {author} {\bibfnamefont {F.~L.~A.}\
  \bibnamefont {Machado}}, \bibinfo {author} {\bibfnamefont {R.~L.}\
  \bibnamefont {{Rodr{\'i}guez-Su{\'a}rez}}}, \bibinfo {author} {\bibfnamefont
  {A.}~\bibnamefont {Azevedo}}, \ and\ \bibinfo {author} {\bibfnamefont
  {S.~M.}\ \bibnamefont {Rezende}},\ }\href {\doibase
  10.1103/PhysRevLett.115.226601} {\bibfield  {journal} {\bibinfo  {journal}
  {Physical Review Letters}\ }\textbf {\bibinfo {volume} {115}},\ \bibinfo
  {pages} {226601} (\bibinfo {year} {2015})}\BibitemShut {NoStop}%
\bibitem [{\citenamefont {Mendes}\ \emph {et~al.}(2019)\citenamefont {Mendes},
  \citenamefont {Alves~Santos}, \citenamefont {Chagas}, \citenamefont
  {{Magalh{\~a}es-Paniago}}, \citenamefont {Mori}, \citenamefont {Holanda},
  \citenamefont {Meireles}, \citenamefont {Lacerda}, \citenamefont {Azevedo},\
  and\ \citenamefont {Rezende}}]{mendesDirectDetectionInduced2019}%
  \BibitemOpen
  \bibfield  {author} {\bibinfo {author} {\bibfnamefont {J.~B.~S.}\
  \bibnamefont {Mendes}}, \bibinfo {author} {\bibfnamefont {O.}~\bibnamefont
  {Alves~Santos}}, \bibinfo {author} {\bibfnamefont {T.}~\bibnamefont
  {Chagas}}, \bibinfo {author} {\bibfnamefont {R.}~\bibnamefont
  {{Magalh{\~a}es-Paniago}}}, \bibinfo {author} {\bibfnamefont {T.~J.~A.}\
  \bibnamefont {Mori}}, \bibinfo {author} {\bibfnamefont {J.}~\bibnamefont
  {Holanda}}, \bibinfo {author} {\bibfnamefont {L.~M.}\ \bibnamefont
  {Meireles}}, \bibinfo {author} {\bibfnamefont {R.~G.}\ \bibnamefont
  {Lacerda}}, \bibinfo {author} {\bibfnamefont {A.}~\bibnamefont {Azevedo}}, \
  and\ \bibinfo {author} {\bibfnamefont {S.~M.}\ \bibnamefont {Rezende}},\
  }\href {\doibase 10.1103/PhysRevB.99.214446} {\bibfield  {journal} {\bibinfo
  {journal} {Physical Review B}\ }\textbf {\bibinfo {volume} {99}},\ \bibinfo
  {pages} {214446} (\bibinfo {year} {2019})}\BibitemShut {NoStop}%
\bibitem [{\citenamefont {Rammer}\ and\ \citenamefont
  {Smith}(1986)}]{rammerQuantumFieldtheoreticalMethods1986}%
  \BibitemOpen
  \bibfield  {author} {\bibinfo {author} {\bibfnamefont {J.}~\bibnamefont
  {Rammer}}\ and\ \bibinfo {author} {\bibfnamefont {H.}~\bibnamefont {Smith}},\
  }\href {\doibase 10.1103/RevModPhys.58.323} {\bibfield  {journal} {\bibinfo
  {journal} {Reviews of Modern Physics}\ }\textbf {\bibinfo {volume} {58}},\
  \bibinfo {pages} {323} (\bibinfo {year} {1986})}\BibitemShut {NoStop}%
\bibitem [{\citenamefont {Adachi}\ \emph {et~al.}(2013)\citenamefont {Adachi},
  \citenamefont {Uchida}, \citenamefont {Saitoh},\ and\ \citenamefont
  {Maekawa}}]{adachiTheorySpinSeebeck2013}%
  \BibitemOpen
  \bibfield  {author} {\bibinfo {author} {\bibfnamefont {H.}~\bibnamefont
  {Adachi}}, \bibinfo {author} {\bibfnamefont {K.-i.}\ \bibnamefont {Uchida}},
  \bibinfo {author} {\bibfnamefont {E.}~\bibnamefont {Saitoh}}, \ and\ \bibinfo
  {author} {\bibfnamefont {S.}~\bibnamefont {Maekawa}},\ }\href {\doibase
  10.1088/0034-4885/76/3/036501} {\bibfield  {journal} {\bibinfo  {journal}
  {Reports on Progress in Physics}\ }\textbf {\bibinfo {volume} {76}},\
  \bibinfo {pages} {036501} (\bibinfo {year} {2013})}\BibitemShut {NoStop}%
\bibitem [{\citenamefont {Ohnuma}\ \emph {et~al.}(2014)\citenamefont {Ohnuma},
  \citenamefont {Adachi}, \citenamefont {Saitoh},\ and\ \citenamefont
  {Maekawa}}]{ohnumaEnhancedDcSpin2014}%
  \BibitemOpen
  \bibfield  {author} {\bibinfo {author} {\bibfnamefont {Y.}~\bibnamefont
  {Ohnuma}}, \bibinfo {author} {\bibfnamefont {H.}~\bibnamefont {Adachi}},
  \bibinfo {author} {\bibfnamefont {E.}~\bibnamefont {Saitoh}}, \ and\ \bibinfo
  {author} {\bibfnamefont {S.}~\bibnamefont {Maekawa}},\ }\href {\doibase
  10.1103/PhysRevB.89.174417} {\bibfield  {journal} {\bibinfo  {journal}
  {Physical Review B}\ }\textbf {\bibinfo {volume} {89}},\ \bibinfo {pages}
  {174417} (\bibinfo {year} {2014})}\BibitemShut {NoStop}%
\bibitem [{\citenamefont {Ohnuma}\ \emph {et~al.}(2017)\citenamefont {Ohnuma},
  \citenamefont {Matsuo},\ and\ \citenamefont
  {Maekawa}}]{ohnumaTheorySpinPeltier2017}%
  \BibitemOpen
  \bibfield  {author} {\bibinfo {author} {\bibfnamefont {Y.}~\bibnamefont
  {Ohnuma}}, \bibinfo {author} {\bibfnamefont {M.}~\bibnamefont {Matsuo}}, \
  and\ \bibinfo {author} {\bibfnamefont {S.}~\bibnamefont {Maekawa}},\ }\href
  {\doibase 10.1103/PhysRevB.96.134412} {\bibfield  {journal} {\bibinfo
  {journal} {Physical Review B}\ }\textbf {\bibinfo {volume} {96}},\ \bibinfo
  {pages} {134412} (\bibinfo {year} {2017})}\BibitemShut {NoStop}%
\bibitem [{\citenamefont {Matsuo}\ \emph {et~al.}(2018)\citenamefont {Matsuo},
  \citenamefont {Ohnuma}, \citenamefont {Kato},\ and\ \citenamefont
  {Maekawa}}]{matsuoSpinCurrentNoise2018}%
  \BibitemOpen
  \bibfield  {author} {\bibinfo {author} {\bibfnamefont {M.}~\bibnamefont
  {Matsuo}}, \bibinfo {author} {\bibfnamefont {Y.}~\bibnamefont {Ohnuma}},
  \bibinfo {author} {\bibfnamefont {T.}~\bibnamefont {Kato}}, \ and\ \bibinfo
  {author} {\bibfnamefont {S.}~\bibnamefont {Maekawa}},\ }\href {\doibase
  10.1103/PhysRevLett.120.037201} {\bibfield  {journal} {\bibinfo  {journal}
  {Physical Review Letters}\ }\textbf {\bibinfo {volume} {120}},\ \bibinfo
  {pages} {037201} (\bibinfo {year} {2018})}\BibitemShut {NoStop}%
\bibitem [{\citenamefont {Inoue}\ \emph {et~al.}(2017)\citenamefont {Inoue},
  \citenamefont {Ichioka},\ and\ \citenamefont
  {Adachi}}]{inoueSpinPumpingSuperconductors2017}%
  \BibitemOpen
  \bibfield  {author} {\bibinfo {author} {\bibfnamefont {M.}~\bibnamefont
  {Inoue}}, \bibinfo {author} {\bibfnamefont {M.}~\bibnamefont {Ichioka}}, \
  and\ \bibinfo {author} {\bibfnamefont {H.}~\bibnamefont {Adachi}},\ }\href
  {\doibase 10.1103/PhysRevB.96.024414} {\bibfield  {journal} {\bibinfo
  {journal} {Physical Review B}\ }\textbf {\bibinfo {volume} {96}},\ \bibinfo
  {pages} {024414} (\bibinfo {year} {2017})}\BibitemShut {NoStop}%
\bibitem [{\citenamefont {Kato}\ \emph {et~al.}(2019)\citenamefont {Kato},
  \citenamefont {Ohnuma}, \citenamefont {Matsuo}, \citenamefont {Rech},
  \citenamefont {Jonckheere},\ and\ \citenamefont
  {Martin}}]{katoMicroscopicTheorySpin2019}%
  \BibitemOpen
  \bibfield  {author} {\bibinfo {author} {\bibfnamefont {T.}~\bibnamefont
  {Kato}}, \bibinfo {author} {\bibfnamefont {Y.}~\bibnamefont {Ohnuma}},
  \bibinfo {author} {\bibfnamefont {M.}~\bibnamefont {Matsuo}}, \bibinfo
  {author} {\bibfnamefont {J.}~\bibnamefont {Rech}}, \bibinfo {author}
  {\bibfnamefont {T.}~\bibnamefont {Jonckheere}}, \ and\ \bibinfo {author}
  {\bibfnamefont {T.}~\bibnamefont {Martin}},\ }\href {\doibase
  10.1103/PhysRevB.99.144411} {\bibfield  {journal} {\bibinfo  {journal}
  {Physical Review B}\ }\textbf {\bibinfo {volume} {99}},\ \bibinfo {pages}
  {144411} (\bibinfo {year} {2019})}\BibitemShut {NoStop}%
\bibitem [{\citenamefont {Nakata}\ \emph {et~al.}(2018)\citenamefont {Nakata},
  \citenamefont {Ohnuma},\ and\ \citenamefont
  {Matsuo}}]{nakataMagnonicNoiseWiedemannFranz2018}%
  \BibitemOpen
  \bibfield  {author} {\bibinfo {author} {\bibfnamefont {K.}~\bibnamefont
  {Nakata}}, \bibinfo {author} {\bibfnamefont {Y.}~\bibnamefont {Ohnuma}}, \
  and\ \bibinfo {author} {\bibfnamefont {M.}~\bibnamefont {Matsuo}},\ }\href
  {\doibase 10.1103/PhysRevB.98.094430} {\bibfield  {journal} {\bibinfo
  {journal} {Physical Review B}\ }\textbf {\bibinfo {volume} {98}},\ \bibinfo
  {pages} {094430} (\bibinfo {year} {2018})}\BibitemShut {NoStop}%
\bibitem [{\citenamefont {Nakata}\ \emph {et~al.}(2019)\citenamefont {Nakata},
  \citenamefont {Ohnuma},\ and\ \citenamefont
  {Matsuo}}]{nakataAsymmetricQuantumShot2019}%
  \BibitemOpen
  \bibfield  {author} {\bibinfo {author} {\bibfnamefont {K.}~\bibnamefont
  {Nakata}}, \bibinfo {author} {\bibfnamefont {Y.}~\bibnamefont {Ohnuma}}, \
  and\ \bibinfo {author} {\bibfnamefont {M.}~\bibnamefont {Matsuo}},\ }\href
  {\doibase 10.1103/PhysRevB.99.134403} {\bibfield  {journal} {\bibinfo
  {journal} {Physical Review B}\ }\textbf {\bibinfo {volume} {99}},\ \bibinfo
  {pages} {134403} (\bibinfo {year} {2019})}\BibitemShut {NoStop}%
\bibitem [{\citenamefont {{\v S}im{\'a}nek}\ and\ \citenamefont
  {Heinrich}(2003)}]{simanekGilbertDampingMagnetic2003}%
  \BibitemOpen
  \bibfield  {author} {\bibinfo {author} {\bibfnamefont {E.}~\bibnamefont {{\v
  S}im{\'a}nek}}\ and\ \bibinfo {author} {\bibfnamefont {B.}~\bibnamefont
  {Heinrich}},\ }\href {\doibase 10.1103/PhysRevB.67.144418} {\bibfield
  {journal} {\bibinfo  {journal} {Physical Review B}\ }\textbf {\bibinfo
  {volume} {67}},\ \bibinfo {pages} {144418} (\bibinfo {year}
  {2003})}\BibitemShut {NoStop}%
\bibitem [{\citenamefont {Tatara}\ and\ \citenamefont
  {Mizukami}(2017)}]{tataraConsistentMicroscopicAnalysis2017}%
  \BibitemOpen
  \bibfield  {author} {\bibinfo {author} {\bibfnamefont {G.}~\bibnamefont
  {Tatara}}\ and\ \bibinfo {author} {\bibfnamefont {S.}~\bibnamefont
  {Mizukami}},\ }\href {\doibase 10.1103/PhysRevB.96.064423} {\bibfield
  {journal} {\bibinfo  {journal} {Physical Review B}\ }\textbf {\bibinfo
  {volume} {96}},\ \bibinfo {pages} {064423} (\bibinfo {year}
  {2017})}\BibitemShut {NoStop}%
\bibitem [{\citenamefont {Novoselov}\ \emph {et~al.}(2005)\citenamefont
  {Novoselov}, \citenamefont {Geim}, \citenamefont {Morozov}, \citenamefont
  {Jiang}, \citenamefont {Katsnelson}, \citenamefont {Grigorieva},
  \citenamefont {Dubonos},\ and\ \citenamefont
  {Firsov}}]{novoselovTwodimensionalGasMassless2005}%
  \BibitemOpen
  \bibfield  {author} {\bibinfo {author} {\bibfnamefont {K.~S.}\ \bibnamefont
  {Novoselov}}, \bibinfo {author} {\bibfnamefont {A.~K.}\ \bibnamefont {Geim}},
  \bibinfo {author} {\bibfnamefont {S.~V.}\ \bibnamefont {Morozov}}, \bibinfo
  {author} {\bibfnamefont {D.}~\bibnamefont {Jiang}}, \bibinfo {author}
  {\bibfnamefont {M.~I.}\ \bibnamefont {Katsnelson}}, \bibinfo {author}
  {\bibfnamefont {I.~V.}\ \bibnamefont {Grigorieva}}, \bibinfo {author}
  {\bibfnamefont {S.~V.}\ \bibnamefont {Dubonos}}, \ and\ \bibinfo {author}
  {\bibfnamefont {A.~A.}\ \bibnamefont {Firsov}},\ }\href {\doibase
  10.1038/nature04233} {\bibfield  {journal} {\bibinfo  {journal} {Nature}\
  }\textbf {\bibinfo {volume} {438}},\ \bibinfo {pages} {197} (\bibinfo {year}
  {2005})}\BibitemShut {NoStop}%
\bibitem [{\citenamefont {Zhang}\ \emph {et~al.}(2005)\citenamefont {Zhang},
  \citenamefont {Tan}, \citenamefont {Stormer},\ and\ \citenamefont
  {Kim}}]{zhangExperimentalObservationQuantum2005}%
  \BibitemOpen
  \bibfield  {author} {\bibinfo {author} {\bibfnamefont {Y.}~\bibnamefont
  {Zhang}}, \bibinfo {author} {\bibfnamefont {Y.-W.}\ \bibnamefont {Tan}},
  \bibinfo {author} {\bibfnamefont {H.~L.}\ \bibnamefont {Stormer}}, \ and\
  \bibinfo {author} {\bibfnamefont {P.}~\bibnamefont {Kim}},\ }\href {\doibase
  10.1038/nature04235} {\bibfield  {journal} {\bibinfo  {journal} {Nature}\
  }\textbf {\bibinfo {volume} {438}},\ \bibinfo {pages} {201} (\bibinfo {year}
  {2005})}\BibitemShut {NoStop}%
\bibitem [{\citenamefont {Kasuya}\ and\ \citenamefont
  {LeCraw}(1961)}]{kasuyaRelaxationMechanismsFerromagnetic1961}%
  \BibitemOpen
  \bibfield  {author} {\bibinfo {author} {\bibfnamefont {T.}~\bibnamefont
  {Kasuya}}\ and\ \bibinfo {author} {\bibfnamefont {R.~C.}\ \bibnamefont
  {LeCraw}},\ }\href {\doibase 10.1103/PhysRevLett.6.223} {\bibfield  {journal}
  {\bibinfo  {journal} {Physical Review Letters}\ }\textbf {\bibinfo {volume}
  {6}},\ \bibinfo {pages} {223} (\bibinfo {year} {1961})}\BibitemShut {NoStop}%
\bibitem [{\citenamefont {Cherepanov}\ \emph {et~al.}(1993)\citenamefont
  {Cherepanov}, \citenamefont {Kolokolov},\ and\ \citenamefont
  {L'vov}}]{cherepanovSagaYIGSpectra1993}%
  \BibitemOpen
  \bibfield  {author} {\bibinfo {author} {\bibfnamefont {V.}~\bibnamefont
  {Cherepanov}}, \bibinfo {author} {\bibfnamefont {I.}~\bibnamefont
  {Kolokolov}}, \ and\ \bibinfo {author} {\bibfnamefont {V.}~\bibnamefont
  {L'vov}},\ }\href {\doibase 10.1016/0370-1573(93)90107-O} {\bibfield
  {journal} {\bibinfo  {journal} {Physics Reports}\ }\textbf {\bibinfo {volume}
  {229}},\ \bibinfo {pages} {81} (\bibinfo {year} {1993})}\BibitemShut
  {NoStop}%
\bibitem [{\citenamefont {Jin}\ \emph {et~al.}(2019)\citenamefont {Jin},
  \citenamefont {Wang}, \citenamefont {Lu}, \citenamefont {Li}, \citenamefont
  {He}, \citenamefont {Zhong},\ and\ \citenamefont
  {Zhang}}]{jinTemperatureDependenceSpinwave2019}%
  \BibitemOpen
  \bibfield  {author} {\bibinfo {author} {\bibfnamefont {L.}~\bibnamefont
  {Jin}}, \bibinfo {author} {\bibfnamefont {Y.}~\bibnamefont {Wang}}, \bibinfo
  {author} {\bibfnamefont {G.}~\bibnamefont {Lu}}, \bibinfo {author}
  {\bibfnamefont {J.}~\bibnamefont {Li}}, \bibinfo {author} {\bibfnamefont
  {Y.}~\bibnamefont {He}}, \bibinfo {author} {\bibfnamefont {Z.}~\bibnamefont
  {Zhong}}, \ and\ \bibinfo {author} {\bibfnamefont {H.}~\bibnamefont
  {Zhang}},\ }\href {\doibase 10.1063/1.5085922} {\bibfield  {journal}
  {\bibinfo  {journal} {AIP Advances}\ }\textbf {\bibinfo {volume} {9}},\
  \bibinfo {pages} {025301} (\bibinfo {year} {2019})}\BibitemShut {NoStop}%
\bibitem [{\citenamefont {Ponomarenko}\ \emph {et~al.}(2010)\citenamefont
  {Ponomarenko}, \citenamefont {Yang}, \citenamefont {Gorbachev}, \citenamefont
  {Blake}, \citenamefont {Mayorov}, \citenamefont {Novoselov}, \citenamefont
  {Katsnelson},\ and\ \citenamefont {Geim}}]{ponomarenkoDensityStatesZero2010}%
  \BibitemOpen
  \bibfield  {author} {\bibinfo {author} {\bibfnamefont {L.~A.}\ \bibnamefont
  {Ponomarenko}}, \bibinfo {author} {\bibfnamefont {R.}~\bibnamefont {Yang}},
  \bibinfo {author} {\bibfnamefont {R.~V.}\ \bibnamefont {Gorbachev}}, \bibinfo
  {author} {\bibfnamefont {P.}~\bibnamefont {Blake}}, \bibinfo {author}
  {\bibfnamefont {A.~S.}\ \bibnamefont {Mayorov}}, \bibinfo {author}
  {\bibfnamefont {K.~S.}\ \bibnamefont {Novoselov}}, \bibinfo {author}
  {\bibfnamefont {M.~I.}\ \bibnamefont {Katsnelson}}, \ and\ \bibinfo {author}
  {\bibfnamefont {A.~K.}\ \bibnamefont {Geim}},\ }\href {\doibase
  10.1103/PhysRevLett.105.136801} {\bibfield  {journal} {\bibinfo  {journal}
  {Physical Review Letters}\ }\textbf {\bibinfo {volume} {105}},\ \bibinfo
  {pages} {136801} (\bibinfo {year} {2010})}\BibitemShut {NoStop}%
\bibitem [{\citenamefont {Banszerus}\ \emph {et~al.}(2015)\citenamefont
  {Banszerus}, \citenamefont {Schmitz}, \citenamefont {Engels}, \citenamefont
  {Dauber}, \citenamefont {Oellers}, \citenamefont {Haupt}, \citenamefont
  {Watanabe}, \citenamefont {Taniguchi}, \citenamefont {Beschoten},\ and\
  \citenamefont {Stampfer}}]{banszerusUltrahighmobilityGrapheneDevices2015}%
  \BibitemOpen
  \bibfield  {author} {\bibinfo {author} {\bibfnamefont {L.}~\bibnamefont
  {Banszerus}}, \bibinfo {author} {\bibfnamefont {M.}~\bibnamefont {Schmitz}},
  \bibinfo {author} {\bibfnamefont {S.}~\bibnamefont {Engels}}, \bibinfo
  {author} {\bibfnamefont {J.}~\bibnamefont {Dauber}}, \bibinfo {author}
  {\bibfnamefont {M.}~\bibnamefont {Oellers}}, \bibinfo {author} {\bibfnamefont
  {F.}~\bibnamefont {Haupt}}, \bibinfo {author} {\bibfnamefont
  {K.}~\bibnamefont {Watanabe}}, \bibinfo {author} {\bibfnamefont
  {T.}~\bibnamefont {Taniguchi}}, \bibinfo {author} {\bibfnamefont
  {B.}~\bibnamefont {Beschoten}}, \ and\ \bibinfo {author} {\bibfnamefont
  {C.}~\bibnamefont {Stampfer}},\ }\href {\doibase 10.1126/sciadv.1500222}
  {\bibfield  {journal} {\bibinfo  {journal} {Science Advances}\ }\textbf
  {\bibinfo {volume} {1}},\ \bibinfo {pages} {e1500222} (\bibinfo {year}
  {2015})}\BibitemShut {NoStop}%
\bibitem [{\citenamefont {Ohshima}\ \emph {et~al.}(2017)\citenamefont
  {Ohshima}, \citenamefont {Klingler}, \citenamefont {Dushenko}, \citenamefont
  {Ando}, \citenamefont {Weiler}, \citenamefont {Huebl}, \citenamefont
  {Shinjo}, \citenamefont {Goennenwein},\ and\ \citenamefont
  {Shiraishi}}]{ohshimaSpinInjectionSilicon2017}%
  \BibitemOpen
  \bibfield  {author} {\bibinfo {author} {\bibfnamefont {R.}~\bibnamefont
  {Ohshima}}, \bibinfo {author} {\bibfnamefont {S.}~\bibnamefont {Klingler}},
  \bibinfo {author} {\bibfnamefont {S.}~\bibnamefont {Dushenko}}, \bibinfo
  {author} {\bibfnamefont {Y.}~\bibnamefont {Ando}}, \bibinfo {author}
  {\bibfnamefont {M.}~\bibnamefont {Weiler}}, \bibinfo {author} {\bibfnamefont
  {H.}~\bibnamefont {Huebl}}, \bibinfo {author} {\bibfnamefont
  {T.}~\bibnamefont {Shinjo}}, \bibinfo {author} {\bibfnamefont {S.~T.~B.}\
  \bibnamefont {Goennenwein}}, \ and\ \bibinfo {author} {\bibfnamefont
  {M.}~\bibnamefont {Shiraishi}},\ }\href {\doibase 10.1063/1.4983012}
  {\bibfield  {journal} {\bibinfo  {journal} {Applied Physics Letters}\
  }\textbf {\bibinfo {volume} {110}},\ \bibinfo {pages} {182402} (\bibinfo
  {year} {2017})}\BibitemShut {NoStop}%
\end{thebibliography}%

\end{document}